\documentclass[%
 reprint,
 superscriptaddress,
 nofootinbib,
 amsmath,amssymb,
 aps,
 prx,
floatfix,
]{revtex4-2}
\pdfoutput=1
\usepackage[T1]{fontenc}
\usepackage[english]{babel}
\usepackage{listings}
\usepackage{xcolor}
\usepackage{amsmath}
\usepackage{graphicx}
\usepackage{adjustbox}
\usepackage{dsfont}
\usepackage{enumitem}
\usepackage{tikz}
\usetikzlibrary{arrows.meta,positioning,fit,calc,shapes.geometric}
\usepackage{amsthm}
\usepackage{tabularx}
\usepackage[caption=false]{subfig}
\usepackage{booktabs}
\usepackage{siunitx}
\usepackage{multirow}
\usepackage{tabularx}
\usepackage{array}
\usepackage{booktabs}
\usepackage{makecell}
\usepackage{multirow}
\usepackage{threeparttable}
\usepackage{array}
\usepackage{xcolor}
\usepackage{colortbl}
\usepackage{pgfplots}
\usepackage{pgfplotstable}
\pgfplotsset{compat=1.18}
\makeatletter
\def\@bibdataout@aps{%
 \immediate\write\@bibdataout{%
  @CONTROL{%
   apsrev42Control%
   \longbibliography@sw{%
    ,author="48",editor="1",pages="0",title="0",year="1"%
   }{%
    ,author="48",editor="1",pages="0",title="",year="1"%
   }%
  }%
 }%
 \if@filesw
  \immediate\write\@auxout{\string\citation{apsrev42Control}}%
 \fi
}%
\let\GigaEvo@PackageWarning\PackageWarning
\def\PackageWarning#1#2{%
 \def\GigaEvo@warningPackage{#1}%
 \def\GigaEvo@namerefPackage{nameref}%
 \ifx\GigaEvo@warningPackage\GigaEvo@namerefPackage
 \else
  \GigaEvo@PackageWarning{#1}{#2}%
 \fi
}%
\makeatother
\usepackage{hyperref}
\makeatletter
\let\PackageWarning\GigaEvo@PackageWarning
\makeatother
\hypersetup{hidelinks}
\definecolor{VGEHeader}{HTML}{DCEAF4}
\definecolor{VGEBody}{HTML}{F2F7FA}
\definecolor{VGEGain}{HTML}{1E6B4E}
\definecolor{VGELoss}{HTML}{8B3A3A}

\newcommand{\vtbest}[1]{\textbf{#1}}
\newcommand{\vtnewbest}[1]{\textbf{#1}\textsuperscript{\(\dagger\)}}
\newcommand{\vtrust}[1]{#1\textsuperscript{\(*\)}}
\newcommand{\vtgain}[2]{#1\,{\scriptsize\textcolor{VGEGain}{\(\downarrow\,#2\)}}}

\newcommand{\vtna}{\textemdash}

\providecommand{\BenchCite}[1]{\cite{#1}}

\newcommand{\ket}[1]{\left|#1\right\rangle}

\begin{document}

\title{LLM-Guided Evolutionary Search for Algebraic \texorpdfstring{$T$}{T}-Count Optimization}

\author{Daniil Fisher}
\email{fisher.de19@physics.msu.ru}
\affiliation{%
Sber Quantum Technologies Center, Moscow, Russia
}
\affiliation{%
 Faculty of Physics, M.V. Lomonosov Moscow State University, Moscow, Russia
}
\author{Valentin Khrulkov}
\affiliation{%
Artificial Intelligence Research Institute (AIRI), Moscow, Russia
}
\author{Mikhail Saygin}
\affiliation{%
Sber Quantum Technologies Center, Moscow, Russia
}
\affiliation{%
 Faculty of Physics, M.V. Lomonosov Moscow State University, Moscow, Russia
}
\author{Ivan Oseledets}
\affiliation{%
Artificial Intelligence Research Institute (AIRI), Moscow, Russia
}
\affiliation{Skolkovo Institute of Science and Technology, Moscow, Russia}
\affiliation{Faculty of Artificial Intelligence, M.V. Lomonosov Moscow State University, Moscow, Russia}
\author{Stanislav Straupe}
\affiliation{%
Sber Quantum Technologies Center, Moscow, Russia
}
\affiliation{%
 Faculty of Physics, M.V. Lomonosov Moscow State University, Moscow, Russia
}
\begin{abstract}
$T$-count minimization is an $\mathrm{NP}$-hard problem that arises in fault-tolerant quantum compilation. In the parity-matrix representation, which captures the non-Clifford part of a quantum circuit, algebraic optimizers such as TODD can achieve state-of-the-art results. However, heuristics fixed in advance determine which transformation is applied, limiting the exploration of alternative trajectories that may lead to better solutions. We show how LLM-guided evolutionary search can help explore these degrees of freedom, which VarTODD exposes through a policy that determines how to allocate the available evaluation budget and how to guide the search. Execution diagnostics guide LLM-generated revisions to both numerical search parameters and program logic, with the possibility of exploiting earlier results by starting from intermediate matrices saved during previous runs. This formulation turns heuristic design into an automated search problem and, across all evaluated instances, matches or improves on the lowest listed reference $T$-count for every evaluated instance; for example, on the $\mathrm{GF}(2^n)$ multiplier benchmarks, it yields a mean relative reduction of $7.0\%$. 
\end{abstract}

\maketitle

\section{Introduction}
\label{sec:Intro}

Fault-tolerant quantum computing requires algorithms to be compiled into gate sets supported by quantum error correction, with Clifford+$T$ providing a widely used gate set. Not all gates are equally expensive to implement: the non-Clifford $T$ gate is often much more expensive than a logical Clifford operation. For example, in a magic-state-based implementation, a $T$ gate requires the preparation and delivery of a high-fidelity non-Clifford resource state~\cite{Fowler2012SurfaceCodes,Litinski2019MagicStateDistillation,GidneyEkera2021RSA2048}. The $T$-count is therefore a standard first-order proxy for logical resource overhead, although $T$-count alone does not determine runtime or physical qubit cost.

Arithmetic circuits are a natural target for this type of optimization because many quantum algorithms rely on the coherent evaluation of classical functions on superpositions. These oracles often reduce to integer, modular, or finite-field arithmetic~\cite{VedralBarencoEkert1996Arithmetic,HanerRoettelerSvore2018Arithmetic}, where modular exponentiation, multiplication, and related arithmetic blocks account for much of the logical non-Clifford overhead~\cite{VedralBarencoEkert1996Arithmetic,GidneyEkera2021RSA2048}. Arithmetic benchmarks therefore combine practical relevance with algebraic structure that exposes the strengths and weaknesses of $T$-count optimizers.

One route to $T$-count minimization is to isolate Hadamard-free or diagonal subproblems, either through Hadamard gadgetization with ancilla qubits or by partitioning circuits into diagonal components~\cite{todd,rm_optimizer,FastTodd,zhang_optimizing_2019}. Such components admit the formalism of phase-polynomials and parity matrices, giving a well-defined mathematical framework for representing the non-Clifford part of the circuit. In this formalism, $T$-count minimization becomes the task of reducing the number of columns in the parity matrix.~\cite{todd,FastTodd,deBeaudrap2020SpiderNest,Ruiz2025AlphaTensorQuantum} and can be naturally linked to coding-theoretic and tensor-decomposition viewpoints~\cite{rm_optimizer,todd,FastTodd}.

Among parity-matrix optimizers, the Third Order Duplicate and Destroy (TODD) algorithm is one of the strongest classical heuristics for iterative $T$-count reduction~\cite{todd}. Its later refinements, including FastTODD and Third Order Homogeneous Polynomials Elimination (TOHPE), substantially improve runtime and match or exceed the quality of earlier approaches~\cite{FastTodd}. Complementary methods include exact and meet-in-the-middle synthesis~\cite{AmyMaslovMoscaRoetteler2013MITM,Gosset2014AlgorithmTCount}, matroid-based $T$-depth optimization~\cite{AmyMaslovMosca2013Matroid}, Pauli-rotation and ZX-calculus rewrites~\cite{zhang_optimizing_2019,KissingerWetering2020PRA,deBeaudrap2020SpiderNest}, and learned or optimized tensor decompositions~\cite{Ruiz2025AlphaTensorQuantum,khoruzhii2026tensordecompositionnoncliffordgate}. Together these methods provide powerful algebraic ingredients, but their practical performance also depends on implementation choices.

At a given parity matrix, TODD and FastTODD may admit many valid candidate actions, and different choices can lead to different terminal matrices, so implementation must decide which candidates to generate and retain, how to score them, and which action to apply next. In standard implementations, these choices are fixed in advance to favor large immediate reductions. We therefore ask whether the policy itself can be optimized while the admissible algebraic updates remain fixed.

VarTODD exposes these decisions as a programmable policy while keeping the FastTODD and TOHPE update rules unchanged. The open question is how to explore this policy space. To explore it, we combine VarTODD with \textsc{GigaEvo}, an evolutionary framework for program synthesis guided by a large language model (LLM)~\cite{GigaEvo}, to evolve the tuner program itself. Candidate programs may define the policy parameterization, numerical search method, restarts and strategy for reusing trajectories, but they cannot bypass an external algebraic validator. This separation combines an auditable algebraic backend with automated search over the surrounding heuristic choices.

The main contributions of this work are as follows.
\begin{enumerate}[leftmargin=*]
    \item We introduce VarTODD, a parametrized extension of FastTODD/TOHPE that exposes the heuristic choices governing the search.
    \item We integrate VarTODD with \textsc{GigaEvo} in a three-level search pipeline that evolves a dedicated tuner for each problem instance, i.e., for each parity matrix representing a quantum circuit.
    \item We demonstrate that the combined VarTODD+\textsc{GigaEvo} pipeline matches or improves upon the lowest listed reference $T$-count across all evaluated benchmarks.
\end{enumerate}

Section~\ref{sec:preliminaries} reviews the parity-matrix representation and identifies the heuristic choices exposed by VarTODD\@. Section~\ref{sec:gigaevo} describes the nested pipeline for evolving tuners and searching policies. Section~\ref{sec:res} reports the benchmark results and computational budgets. Section~\ref{sec:discussion} discusses the scope of the $T$-count objective and positions program evolution relative to learned search policies. Section~\ref{sec:conclusion} summarizes the main findings and limitations.

\section{Preliminaries}
\label{sec:preliminaries}
This section reviews the basic concepts of quantum computation, defines the parity-matrix optimization problem, and introduces the search controls exposed by VarTODD. Appendix~\ref{app:parity_background} provides further details on the signature tensor and the update rules used by TODD, FastTODD, and TOHPE.

\subsection{Clifford+\texorpdfstring{$T$}{T} computation}
\label{subsec:physical_background}

An $n$-qubit pure quantum state is represented by a unit vector in $\mathcal{H}_n=(\mathbb{C}^{2})^{\otimes n}$, and a quantum circuit implements a unitary transformation on this space~\cite{NielsenChuang2010}.

The Clifford group is the normalizer of the $n$-qubit Pauli group,
\begin{equation}
    \mathcal{C}_n = \{U:U\mathcal{P}_nU^{\dagger}=\mathcal{P}_n\}.
    \label{eq:clifford_group}
\end{equation}
Up to global phase, it is generated by the single-qubit Hadamard gate $H$, the phase gate $S$, and the two-qubit CNOT gate and admit efficient classical simulation~\cite{Gottesman1998Heisenberg}. Adding the non-Clifford $T$ gate, which applies the phase $e^{i\pi/4}$ to the state $\ket{1}$, yields the approximately universal Clifford+$T$ gate set~\cite{NielsenChuang2010,RossSelinger2016}.

The Clifford/non-Clifford distinction is operationally important in fault-tolerant architectures. The Eastin--Knill theorem precludes a universal set of transversal logical gates for a fixed finite-dimensional quantum code, so universal schemes require additional mechanisms such as magic-state injection, code switching, gauge fixing, or pieceable fault-tolerant constructions~\cite{EastinKnill2009,BravyiKitaev2005,PaetznickReichardt2013,Bombin2015GaugeColorCodes,YoderTakagiChuang2016}. The precise overhead is architecture dependent, but logical non-Clifford operations are commonly much more demanding than logical Clifford operations.

\subsection{Phase polynomials and parity matrices}
\label{subsec:tcount_min}
For a Hadamard-free CNOT+$T$ circuit, the unitary operator can be written as
\begin{equation}
U\ket{x} = \omega^{f(x)}\ket{Ax}, \qquad f(x) = \sum_{j=1}^{r} c_j\langle p_j,x\rangle_{\mathbb F_2} \pmod 8,
\end{equation}
where $x\in\mathbb F_2^n$ and $\omega=e^{i\pi/4}$. The matrix $A\in\mathrm{GL}(n,\mathbb F_2)$ describes the reversible linear transformation implemented by the CNOT gates. Each $p_j\in\mathbb F_2^n$ specifies a parity, and $c_j\in\mathbb Z_8$ is its phase coefficient~\cite{rm_optimizer,todd,FastTodd}.

Terms with even coefficients contribute only Clifford phases. After absorbing these terms into the Clifford part and reindexing the remaining $m$ odd-coefficient terms, we collect their parity vectors as the columns of the \emph{parity matrix}
\begin{equation}
P=[p_1\ \cdots\ p_m]\in\mathbb F_2^{n\times m}.
\end{equation} 
Each odd coefficient requires one $T$-type gate up to Clifford corrections, so the number of columns $m$ is the $T$-count of this representation. The parity matrix does not retain the individual odd coefficients and therefore does not distinguish, for example, a $T$ contribution from a $T^\dagger$ contribution.
\begin{figure}[hbtp!]
    \centering
    \begin{tikzpicture}[
        x=0.88cm,
        y=0.88cm,
        every node/.style={font=\small},
        phase/.style={draw, minimum width=0.62cm, minimum height=0.52cm,
                     inner sep=1pt, fill=white},
        paritylabel/.style={font=\scriptsize}
    ]
        \draw (0,1) -- (4.8,1);
        \draw (0,0) -- (4.8,0);
        \node[left] at (0,1) {$\ket{x_1}$};
        \node[left] at (0,0) {$\ket{x_2}$};

        \node[phase] (t1) at (1.0,1) {$T$};
        \node[phase] (t2) at (1.0,0) {$T$};
        \node[paritylabel, above=1pt of t1] {$p_1^{\mathsf T}x=x_1$};
        \node[paritylabel, below=1pt of t2] {$p_2^{\mathsf T}x=x_2$};

        \fill (2,1) circle (1.8pt);
        \draw (2,1) -- (2,0);
        \node at (2,0) {$\oplus$};

        \node[phase] (td) at (3.0,0) {$T^{\dagger}$};
        \node[paritylabel, below=1pt of td]
            {$p_3^{\mathsf T}x=x_1\oplus x_2$};

        \fill (4,1) circle (1.8pt);
        \draw (4,1) -- (4,0);
        \node at (4,0) {$\oplus$};
        \node at (5.4,0.5) {$\longmapsto$};
        \node at (7.1,0.5) {$P=\begin{pmatrix} 1 & 0 & 1\\ 0 & 1 & 1 \end{pmatrix}$};
    \end{tikzpicture}
    \caption{Extraction of the parity matrix for a Clifford+$T$ implementation of the controlled-$S$ gate. Each column of $P$ records the parity carried by the wire at one $T$-type gate. The corresponding phase coefficients are $c=(1,1,7)$, where $7$ represents the $T^{\dagger}$ contribution.}
    \label{fig:cs_parity_example}
\end{figure}
Figure~\ref{fig:cs_parity_example} illustrates the construction for a Clifford+$T$ implementation of the controlled-$S$ gate, which correspond to the phase polynomial
\begin{equation}
f_{\mathrm{CS}}(x) = x_1+x_2+7(x_1\oplus x_2) \equiv 2x_1x_2 \pmod 8.
\end{equation}

To validate a decomposition, we use the symmetric signature tensor associated with a parity matrix $P$, defined by
\begin{equation}
S_{\alpha\beta\gamma}(P) = \sum_{j=1}^{m} P_{\alpha j}P_{\beta j}P_{\gamma j} \pmod 2,
\end{equation}
where repeated indices are allowed. Two parity matrices represent the same non-Clifford phase up to diagonal Clifford corrections if and only if their signature tensors agree~\cite{todd,FastTodd,rm_optimizer,Ruiz2025AlphaTensorQuantum}. Thus, given an initial parity matrix $P_0$, $T$-count minimization can be reduced to finding an equivalent parity matrix with the smallest possible number of columns:
\begin{equation}
P^\star \in
\underset{P':\,S(P')=S(P_0)}{\arg\min}
\operatorname{cols}(P').
\end{equation}

\subsection{Signature-preserving updates and action selection}

TODD, FastTODD, and TOHPE search for updates that preserve the signature tensor while creating duplicate or zero columns. For a current matrix $P$, an admissible action is a pair
\begin{equation}
a=(z,y), \qquad z\in\mathbb F_2^n, \qquad y\in N_z(P)\subseteq\mathbb F_2^m,
\end{equation}
where $N_z(P)$ is the admissible space constructed by the algebraic backend. The update is
\begin{equation}
P'=\operatorname{odd}(P\oplus zy^T),
\end{equation}
where $\operatorname{odd}(\cdot)$ removes zero columns and cancels duplicate columns in pairs. These simplifications preserve the signature tensor and affect only the diagonal Clifford correction, thereby preserving the represented non-Clifford phase.

\begin{figure}
    \centering
    \includegraphics[width=1.0\linewidth]{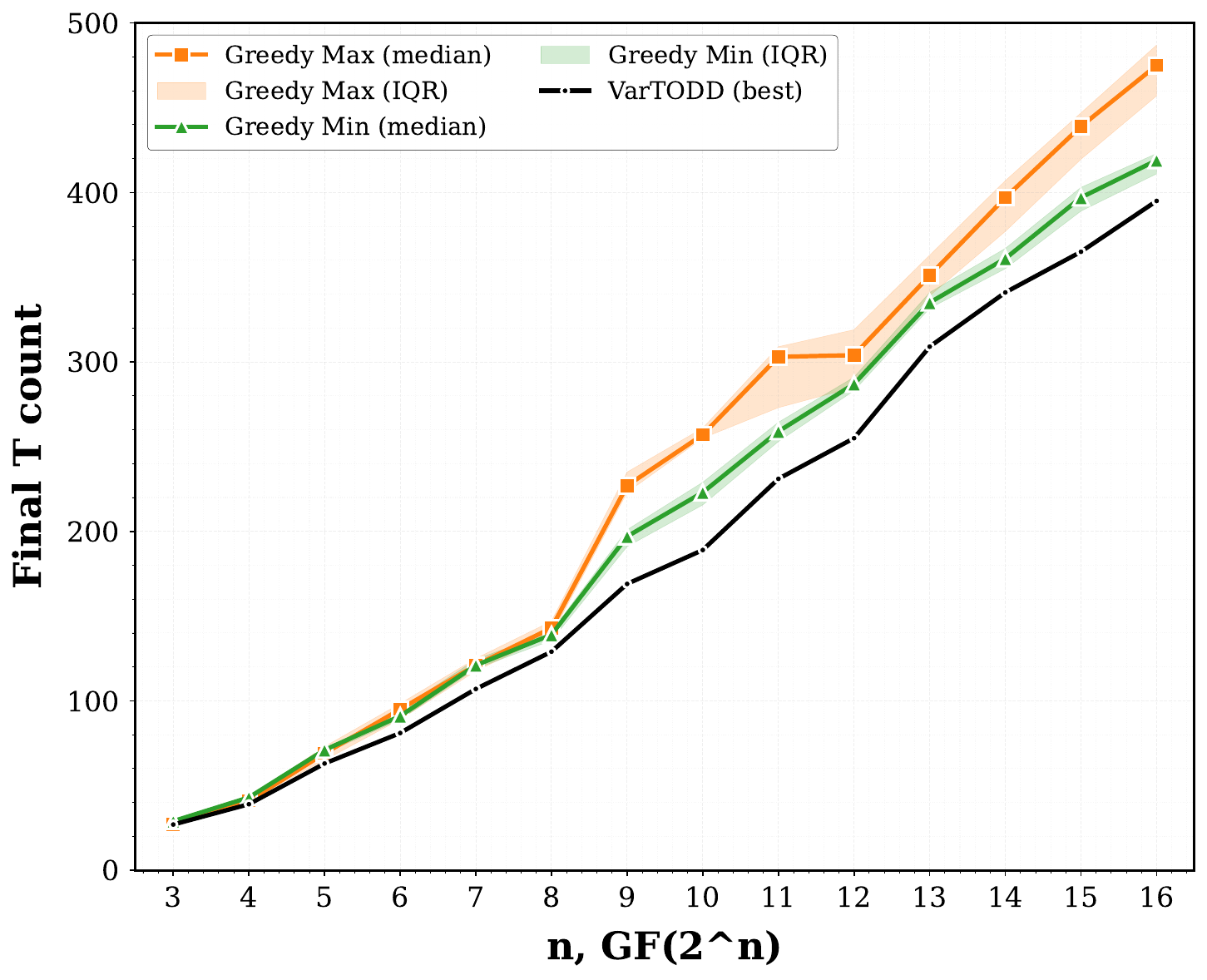}
    \caption{Illustrative comparison of two simple action-selection rules in a parity-matrix optimizer. \emph{Greedy Max} selects the examined action with the largest immediate column-count reduction, whereas \emph{Greedy Min} selects the smallest positive reduction. Solid curves show medians and the shaded regions show interquartile ranges over random seeds. The different trajectories motivate treating action selection as an explicit policy-optimization problem.}
    \label{fig:gf2_greedy_compare}
\end{figure}

The algebraic conditions determine which actions are valid, but they do not determine which valid action should be examined or applied. Different choices can produce different successor matrices and, consequently, different terminal $T$-counts, as illustrated in Fig.~\ref{fig:gf2_greedy_compare}. Even a simple reversal of the immediate-reduction preference, from selecting the largest reduction to selecting the smallest positive one, can lead to a lower terminal $T$-count.

\begin{figure*}
    \centering
    \includegraphics[width=1.0\linewidth]{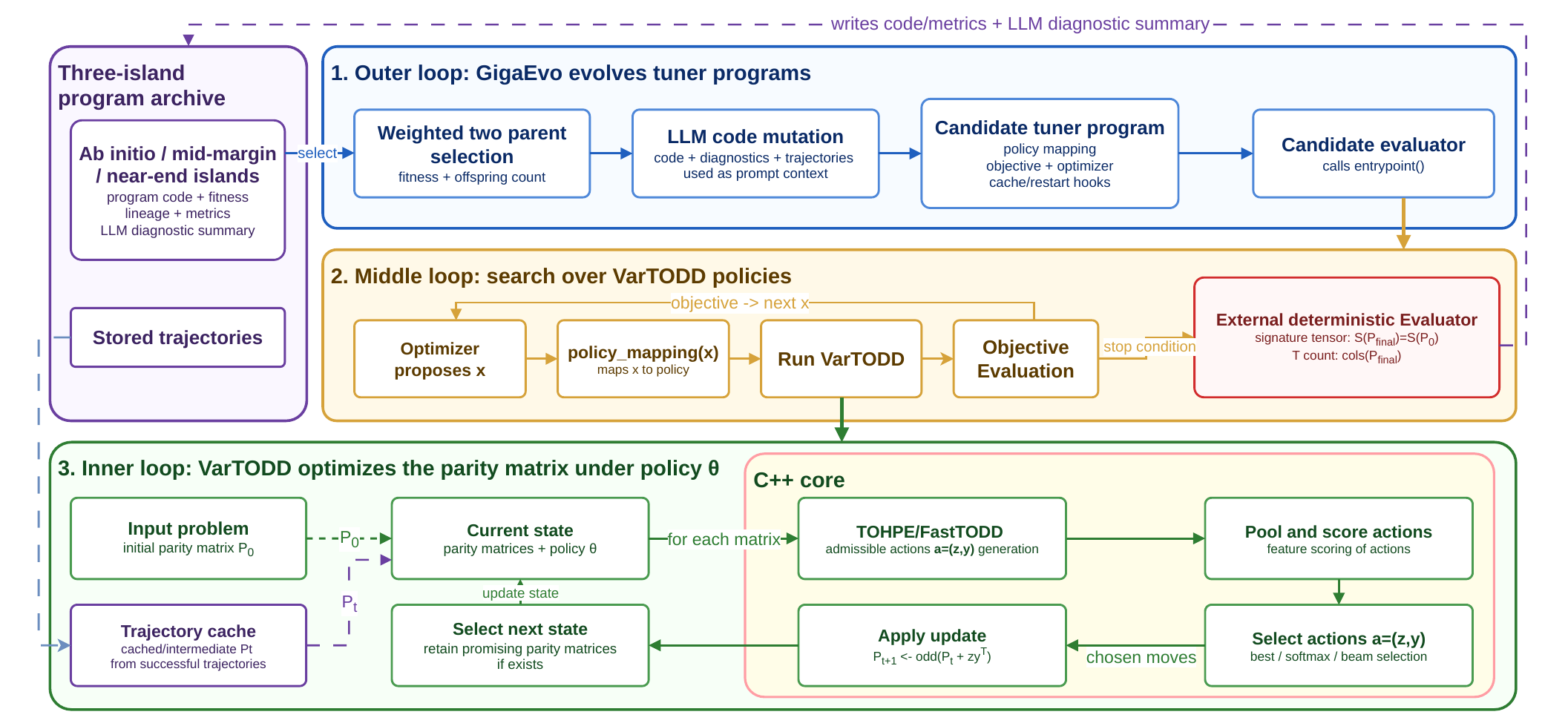}
    \caption{Three-level search pipeline. The outer \textsc{GigaEvo} loop evolves tuner programs from archived code and diagnostics. Each tuner program defines a policy parameterization, numerical search procedure, schedules, and restart logic. The middle loop evaluates policy parameters, and the VarTODD backend performs the admissible descent over parity matrices. An external evaluator verifies equality of the signature tensors and returns the final column count; validated trajectories may be cached for later restarts.}
    \label{fig:gigaevo_vartodd_pipeline}
\end{figure*}

\label{subsec:policy_black_box}
\label{subsec:heuristic_action_selection}

VarTODD keeps the algebraic update rule fixed while exposing the surrounding search decisions as a configurable policy $\theta$. At a given parity matrix $P$, the policy determines how candidate successor matrices with reduced column count are generated, evaluated, and selected. In particular, it distributes the available evaluation budget between the TOHPE and FastTODD candidate generators. Each generator maintains its own candidate pool, retaining only the top-$k$ actions according to its ranking rule. The retained candidates are then compared using a parameterized scoring function. Different policies may therefore favor different types of actions: for example, some may prioritize the largest immediate column-count reduction, resulting in a greedier search, whereas others may favor actions that lead to successor matrices with a larger TOHPE search-space dimension. Once the generator-specific pools have been formed, the policy determines how candidates from the two sources are compared and which retained actions are selected for the next search step. The budget allocation, number of generated samples, pool sizes, scoring parameters, and final selection rule therefore jointly determine how the admissible-action space is explored.

These controls need not remain fixed throughout a VarTODD descent. They may depend on the current $T$-count through predefined schedules, allowing, for example, the relative TOHPE/FastTODD budget, candidate-pool sizes, or scoring parameters to change as the $T$-count decreases. VarTODD can therefore adopt different search behavior in different regions of the optimization trajectory while leaving the underlying algebraic transformations unchanged. Appendix~\ref{app:vartodd} lists the complete set of controls.

Starting from $P_0$, VarTODD repeatedly generates admissible actions, evaluates them according to the current policy, and advances to one or more successor matrices. With beam width one, the search follows a single trajectory, whereas larger beam widths retain several successor states and allow multiple trajectories to be explored in parallel. The search terminates when its evaluation budget is exhausted or when the candidate generators return no reducing action. VarTODD then returns the matrix with the smallest column count encountered during the run.

\section{Automated policy search with \textsc{GigaEvo}}
\label{sec:gigaevo}

We organize the search into three nested levels: the evolution of tuner programs with \textsc{GigaEvo}, tuner-defined search over VarTODD policies, and the inner VarTODD descent. Importantly, the middle level is not restricted to numerical optimization of a fixed parameter vector. A tuner program may define how policies are parameterized, sampled, optimized, and combined, as well as how VarTODD is invoked with different schedules, beam widths, or restart rules. This section describes the complete search pipeline, the three-island archive, parent selection, and the reuse of validated trajectories.


\begin{table*}[t]
  \centering
  \begin{threeparttable}
    \caption{\(\mathrm{GF}(2^n)\) multiplication benchmarks. Entries are final \(T\)-counts; lower is better.}
    \label{tab:gf-mult}
    \small
    \setlength{\tabcolsep}{4.2pt}
    \renewcommand{\arraystretch}{1.10}
    \begin{tabular*}{\textwidth}{@{\extracolsep{\fill}} l r rr c rr >{\columncolor{VGEBody}}c >{\columncolor{VGEBody}}c @{}}
      \toprule
      & \multicolumn{3}{c}{\textbf{Input}}
      & \multicolumn{3}{c}{\textbf{Reference methods}}
      & \multicolumn{2}{c}{\cellcolor{VGEHeader}\textbf{This work}} \\
      \cmidrule(lr){2-4}\cmidrule(lr){5-7}\cmidrule(lr){8-9}
      \multirow{2}{*}{Circuit}
      & \multirow{2}{*}{\makecell{Qubit\\number}}
      & \multicolumn{2}{c}{Initial \(T\)-count}
      & \multirow{2}{*}{\makecell{AlphaTensor\\Quantum}}
      & \multicolumn{2}{c}{FastTODD}
      & \multicolumn{2}{c}{\cellcolor{VGEHeader}\makecell{\textbf{VarTODD +}\\\textsc{\textbf{GigaEvo}}}} \\
      \cmidrule(lr){3-4}\cmidrule(lr){6-7}\cmidrule(lr){8-9}
      & & V & K & & V & K & \cellcolor{VGEHeader}V & \cellcolor{VGEHeader}K \\
      \midrule
      \(\mathrm{GF}(2^3)\)    &  9 &  40 &  42 & 29 & \vtrust{\vtbest{27}} & \vtrust{29} & \vtbest{27} & 29 \\
      \(\mathrm{GF}(2^4)\)    & 12 &  56 &  63 & \vtbest{39} & \vtrust{41} & \vtrust{47} & \vtgain{\vtbest{39}}{2} & \vtgain{\vtbest{39}}{8} \\
      \(\mathrm{GF}(2^5)\)    & 15 &  98 &  91 & 59 & \vtrust{75} & \vtrust{65} & \vtgain{\vtnewbest{49}}{26} & \vtgain{59}{6} \\
      \(\mathrm{GF}(2^6)\)    & 18 & 132 & 105 & \vtbest{77} & \vtrust{93} & \vtrust{81} & \vtgain{79}{14} & \vtgain{\vtbest{77}}{4} \\
      \(\mathrm{GF}(2^7)\)    & 21 & 164 & 150 & 104 & \vtrust{117} & \vtrust{107} & \vtgain{105}{12} & \vtgain{\vtnewbest{101}}{6} \\
      \(\mathrm{GF}(2^8)\)    & 24 & 189 & 168 & \vtbest{123} & \vtrust{141} & \vtrust{129} & \vtgain{\vtbest{123}}{18} & \vtgain{\vtbest{123}}{6} \\
      \(\mathrm{GF}(2^9)\)    & 27 & 320 & 210 & 161 & \vtrust{195} & \vtrust{147} & \vtgain{165}{30} & \vtgain{\vtnewbest{139}}{8} \\
      \(\mathrm{GF}(2^{10})\) & 30 & 334 & 231 & 196 & \vtrust{227} & \vtrust{173} & \vtgain{187}{40} & \vtgain{\vtnewbest{157}}{16} \\
      \(\mathrm{GF}(2^{11})\) & 33 & 376 & \vtna & \vtna & \vtrust{255} & \vtna & \vtgain{\vtnewbest{215}}{40} & \vtna \\
      \(\mathrm{GF}(2^{12})\) & 36 & 396 & \vtna & \vtna & \vtrust{281} & \vtna & \vtgain{\vtnewbest{243}}{38} & \vtna \\
      \(\mathrm{GF}(2^{13})\) & 39 & 460 & \vtna & \vtna & \vtrust{327} & \vtna & \vtgain{\vtnewbest{293}}{34} & \vtna \\
      \(\mathrm{GF}(2^{14})\) & 42 & 492 & \vtna & \vtna & \vtrust{361} & \vtna & \vtgain{\vtnewbest{319}}{42} & \vtna \\
      \(\mathrm{GF}(2^{15})\) & 45 & 536 & \vtna & \vtna & \vtrust{391} & \vtna & \vtgain{\vtnewbest{363}}{28} & \vtna \\
      \(\mathrm{GF}(2^{16})\) & 48 & 567 & \vtna & \vtna & \vtrust{423} & \vtna & \vtgain{\vtnewbest{383}}{40} & \vtna \\
      \(\mathrm{GF}(2^{32})\) & 96 & 1701 & \vtna & \vtna & \vtrust{1275} & \vtna & \vtgain{\vtnewbest{1177}}{98} & \vtna \\
      \(\mathrm{GF}(2^{64})\) & 192 & 5103 & \vtna & \vtna & \vtrust{3815} & \vtna & \vtgain{\vtnewbest{3729}}{86} & \vtna \\
      \bottomrule
    \end{tabular*}
    \begin{tablenotes}[flushleft]
      \footnotesize
      \item[] \textit{Sources and notation.}
      V and K denote the Vandaele and Khoruzhii initial decompositions, respectively
      \BenchCite{FastTodd,GF2MultiplicationVandaele,khoruzhii2026tensordecompositionnoncliffordgate};
      comparisons between initialization-dependent methods should be made within the same branch.
      AlphaTensor Quantum values are reproduced from Ref.~\BenchCite{Ruiz2025AlphaTensorQuantum}.
      An asterisk marks a FastTODD value obtained with Vandaele's Rust implementation on the corresponding initial parity matrix.
      Bold marks the row best, \(\dagger\) a result from this work strictly below every listed reference, and \textemdash\ an unavailable initial decomposition.
      In the shaded columns, \(\downarrow k\) reports a branch-matched reduction of \(k\) relative to FastTODD.
    \end{tablenotes}
  \end{threeparttable}
\end{table*}

\begin{table*}[t]
  \centering
  \begin{threeparttable}
    \caption{Additional  benchmarks. Entries are final \(T\)-counts; lower is better.}
    \label{tab:other-circuits}
    \small
    \setlength{\tabcolsep}{5.0pt}
    \renewcommand{\arraystretch}{1.08}
    \begin{tabular*}{\textwidth}{@{\extracolsep{\fill}} l rr rr >{\columncolor{VGEBody}}c @{}}
      \toprule
      & \multicolumn{2}{c}{\textbf{Input}}
      & \multicolumn{2}{c}{\textbf{Reference methods}}
      & \cellcolor{VGEHeader}\textbf{This work} \\
      \cmidrule(lr){2-3}\cmidrule(lr){4-5}\cmidrule(lr){6-6}
      Circuit
      & \makecell{Qubit\\number}
      & \makecell{Initial\\\(T\)-count}
      & Khoruzhii
      & FastTODD
      & \cellcolor{VGEHeader}\makecell{\textbf{VarTODD +}\\\textsc{\textbf{GigaEvo}}} \\
      \midrule
      Adder 8        &  61 &  173 & \vtbest{117} & \vtrust{119} & \vtgain{\vtbest{117}}{2} \\
      Barenco Tof 3  &   8 &   16 & \vtbest{13}  & \vtrust{\vtbest{13}} & \vtbest{13} \\
      Barenco Tof 4  &  14 &   28 & \vtbest{23}  & \vtrust{\vtbest{23}} & \vtbest{23} \\
      Barenco Tof 5  &  20 &   40 & \vtbest{33}  & \vtrust{\vtbest{33}} & \vtbest{33} \\
      Barenco Tof 10 &  50 &  100 & \vtbest{83}  & \vtrust{\vtbest{83}} & \vtbest{83} \\
      CSLA MUX 3     &  21 &   62 & \vtbest{39}  & \vtrust{\vtbest{39}} & \vtbest{39} \\
      CSUM MUX 9     &  42 &   84 & \vtbest{71}  & \vtrust{\vtbest{71}} & \vtbest{71} \\
      Grover 5       &  77 &  166 & \vtbest{143} & \vtbest{143} & \vtbest{143} \\
      Ham15 (low)    &  97 &  256 & \vtbest{73} & \vtbest{73} & \vtbest{73} \\
      Ham15 (medium) &  71 &  212 & 137 & \vtrust{137} & \vtgain{\vtnewbest{135}}{2} \\
      Ham15 (high)   & 351 & 1019 & 643 & 639 & \vtgain{\vtnewbest{631}}{8} \\
      Mod 5-4        &   5 &    8 & \vtbest{7}   & \vtrust{\vtbest{7}} & \vtbest{7} \\
      Mod Adder 1024 & 331 & 1011 & \vtbest{573} & \vtbest{573} & \vtbest{573} \\
      Mod Mult 5-5   &  11 &   35 & \vtbest{17}  & \vtrust{\vtbest{17}} & \vtbest{17} \\
      Mod Red 2-1    &  28 &   73 & \vtbest{51}  & \vtrust{\vtbest{51}} & \vtbest{51} \\
      QCLA Adder 10  &  61 &  162 & \vtbest{107} & \vtrust{109} & \vtgain{\vtbest{107}}{2} \\
      QCLA Com 7     &  42 &   95 & \vtbest{59}  & \vtbest{59} & \vtbest{59} \\
      QCLA Mod 7     &  84 &  237 & \vtbest{153} & \vtrust{159} & \vtgain{\vtbest{153}}{6} \\
      QFT 4          &  43 &   67 & \vtbest{53}  & \vtbest{53} & \vtbest{53} \\
      RC Adder 6     &  24 &   47 & \vtbest{37}  & \vtrust{\vtbest{37}} & \vtbest{37} \\
      Tof 3          &   7 &   15 & \vtbest{13}  & \vtbest{13} & \vtbest{13} \\
      Tof 4          &  11 &   23 & \vtbest{19}  & \vtbest{19} & \vtbest{19} \\
      Tof 5          &  15 &   31 & \vtbest{25}  & \vtbest{25} & \vtbest{25} \\
      Tof 10         &  35 &   71 & \vtbest{55}  & \vtbest{55} & \vtbest{55} \\
      VBE Adder 3    &  14 &   24 & \vtbest{19}  & \vtbest{19} & \vtbest{19} \\
      \bottomrule
    \end{tabular*}
    \begin{tablenotes}[flushleft]
      \footnotesize
      \item[] \textit{Sources and notation.}
      Khoruzhii values are reproduced from Ref.~\BenchCite{khoruzhii2026tensordecompositionnoncliffordgate};
      FastTODD entries were obtained with Vandaele's Rust implementation on the corresponding initial parity matrices.
      The shaded column reports the best validated VarTODD+\textsc{GigaEvo} result.
      Bold marks the row best and \(\dagger\) a result from this work strictly below every listed reference.
      In the shaded column, \(\downarrow k\) reports a reduction of \(k\) relative to FastTODD.
    \end{tablenotes}
  \end{threeparttable}
\end{table*}
\subsection{Three-level search pipeline}

The complete \textsc{GigaEvo} search pipeline is summarized in Fig.~\ref{fig:gigaevo_vartodd_pipeline}. In the outer loop, \textsc{GigaEvo} maintains an archive of tuner programs divided into three islands and uses an LLM to propose offspring from parent code, task instructions, and diagnostics from previous evaluations. In the reported experiments, proposals were generated using a mixture of DeepSeek V4 Flash and GPT-5 mini.

For each candidate tuner program, the middle loop executes the search procedure defined by that tuner. A common strategy is to map a bounded real-valued vector $x$ to a concrete VarTODD policy $\theta$ and optimize $x$ using a numerical optimization method. However, the interface is not restricted to this parameterized form. A tuner may instead sample policies stochastically, apply one or more numerical optimizers, directly compare several policy configurations, or combine these operations in multiple stages. The evolved program therefore determines not only the values of individual policy parameters, but also the procedure used to search over them.

Each policy considered by the tuner is passed to the inner VarTODD loop, which performs the admissible descent over parity matrices described in Sec.~\ref{subsec:policy_black_box}. The algebraic backend and its admissibility conditions remain fixed regardless of the search procedure chosen by the tuner.

The resulting parity matrices are evaluated separately from the generated tuner program. For every reported output $P_{\mathrm{final}}$, an external evaluator independently recomputes the signature tensor and requires
\begin{equation}
S(P_{\mathrm{final}})=S(P_0).
\label{eq:external_validation}
\end{equation}
Only matrices that satisfy Eq.~\eqref{eq:external_validation} are accepted, and the fitness of a tuner program is given by the smallest column count among its validated outputs. Generated code cannot alter or bypass the criterion used to verify correctness.

\subsection{Three-island archive, parent selection, and trajectory reuse}
\label{subsec:gigaevo_eval_loop}

The outer search is organized around a three-island MAP-Elites archive~\cite{MAPElites}, separating ab initio exploration from two forms of trajectory reuse. This organization balances exploration and exploitation during the search.

The first island, which we refer to as \emph{ab initio}, contains tuner programs that always begin with the original parity matrix $P_0$. \emph{Mid-margin} programs restart from validated intermediate matrices, avoiding repeated evaluation of the early part of a trajectory, whereas \emph{near-end} programs restart close to the endpoint of strong trajectories and focus computation on the final reductions. To generate an offspring, GigaEvo selects two archive members with probabilities that increase as their validated $T$-counts decrease and decrease with offspring count. Their code and execution summaries are then supplied to the LLM, which proposes a modified tuner. This organization preserves full-run exploration while allowing later generations to exploit validated partial trajectories.

\section{Results}
\label{sec:res}

We compare the combined VarTODD+\textsc{GigaEvo} pipeline with published and reproduced reference methods on two benchmark families. Table~\ref{tab:gf-mult} covers $\mathrm{GF}(2^n)$ multiplication circuits, whose non-Clifford components can be extracted exactly into a representation by phase polynomials and parity matrices without gadgetization. Table~\ref{tab:other-circuits} reports selected additional benchmarks from the literature on non-Clifford optimization~\cite{Amy2016Feynman,Ruiz2025AlphaTensorQuantum,khoruzhii2026tensordecompositionnoncliffordgate}. Because different circuit decompositions can produce different initial parity matrices, comparisons are made within the corresponding initialization branch.

Across the multiplication suite, VarTODD+\textsc{GigaEvo} obtains a T-count below the lowest listed reference value for 12 of the 16 tested instances, as shown in Table~\ref{tab:gf-mult}. It matches the corresponding value for $\mathrm{GF}(2^3)$, $\mathrm{GF}(2^4)$, $\mathrm{GF}(2^6)$, and $\mathrm{GF}(2^8)$, yielding a mean relative $T$-count reduction of 7.0\% across the complete suite. On the additional  benchmarks, the pipeline reduces Ham15 (medium) from 137 to 135 and Ham15 (high) from 639 to 631, while the remaining results in Table~\ref{tab:other-circuits} equal the strongest listed values.

\subsection{Experimental setup and parallel execution}
\begin{figure*}[t]
    \centering
    \subfloat[GF$(2^{10})$]{%
        \includegraphics[width=0.32\textwidth]{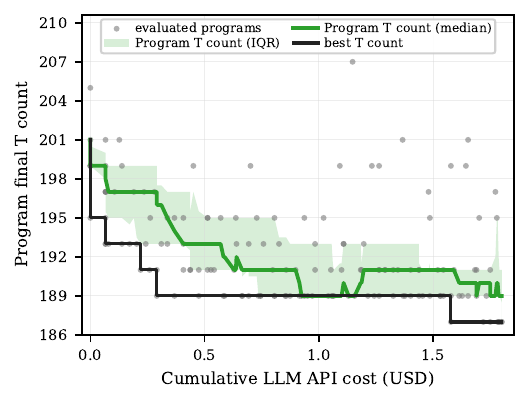}%
        \label{fig:gigaevo_cost_gf10}
        }
    \hfill
    \subfloat[GF$(2^{11})$]{%
        \includegraphics[width=0.32\textwidth]{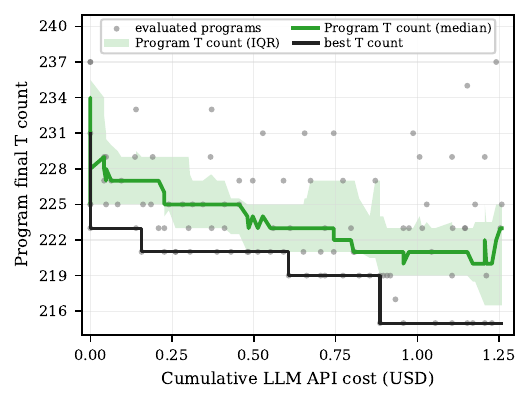}%
        \label{fig:gigaevo_cost_gf11}
    }
    \hfill
    \subfloat[GF$(2^{12})$]{%
        \includegraphics[width=0.32\textwidth]{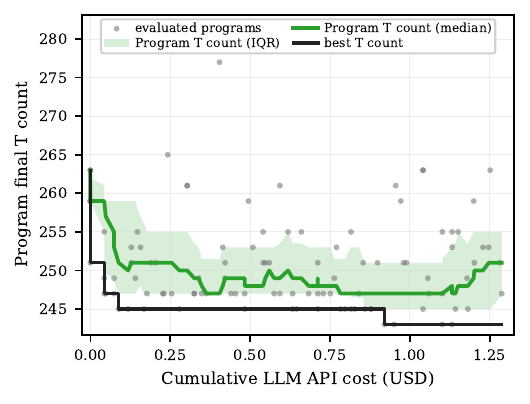}%
        \label{fig:gigaevo_cost_gf12}
        }
    \par\medskip
    \subfloat[GF$(2^{13})$]{%
        \includegraphics[width=0.32\textwidth]{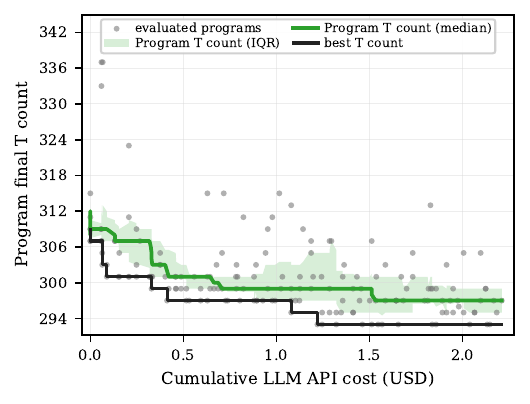}%
        \label{fig:gigaevo_cost_gf13}
        }
    \hfill
    \subfloat[GF$(2^{14})$]{%
        \includegraphics[width=0.32\textwidth]{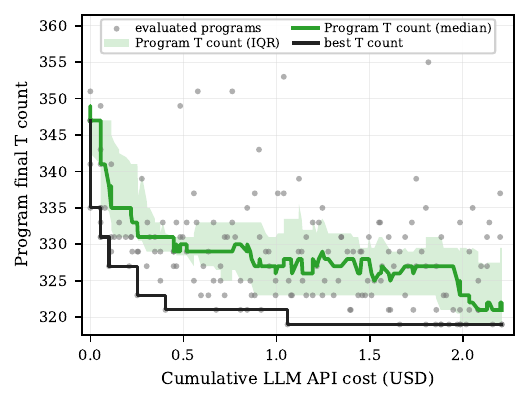}%
        \label{fig:gigaevo_cost_gf14}
        }
    \hfill
    \subfloat[GF$(2^{15})$]{%
        \includegraphics[width=0.32\textwidth]{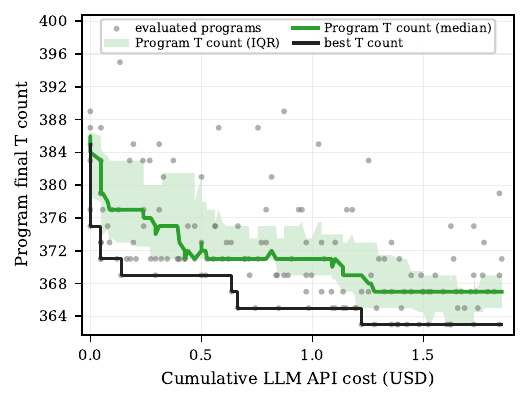}%
        \label{fig:gigaevo_cost_gf15}
    }
    \par\medskip
    \caption{Cost diagnostics for the VarTODD+\textsc{GigaEvo} pipeline on selected $\mathrm{GF}(2^n)$ multiplication instances. Each point represents one evaluated tuner program and shows the smallest validated final column count returned by its inner VarTODD search. The rolling-median curve and shaded band show the median and interquartile range over a window of 28 consecutive program evaluations. The step curve shows the incumbent best validated column count as a function of cumulative LLM API cost for the mixture of DeepSeek V4 Flash and GPT-5 mini used in the experiments.}
    \label{fig:gigaevo_cost_diagnostics}
\end{figure*}

For most benchmark circuits, we performed one instance-specific outer-evolution run and report the smallest validated $T$-count obtained. For $\mathrm{GF}(2^{16})$, we conducted seven separate evolution runs with different random seeds and, in some cases, different initial tuner programs. Candidate tuner programs were evaluated in parallel on 12 CPU cores, with up to three hours allotted to each program for $\mathrm{GF}(2^{64})$. Program generation also incurred LLM API costs.

Once generated, candidate tuners can be evaluated largely independently, so their evaluations can be parallelized. Additional workers would increase evaluation throughput and allow more program variants and search trajectories to be examined concurrently without modifying the underlying search procedure. 

\subsection{Search trajectories and variation across runs}

\begin{figure}[hbtp!]
    \centering
    \includegraphics[width=0.9\linewidth]{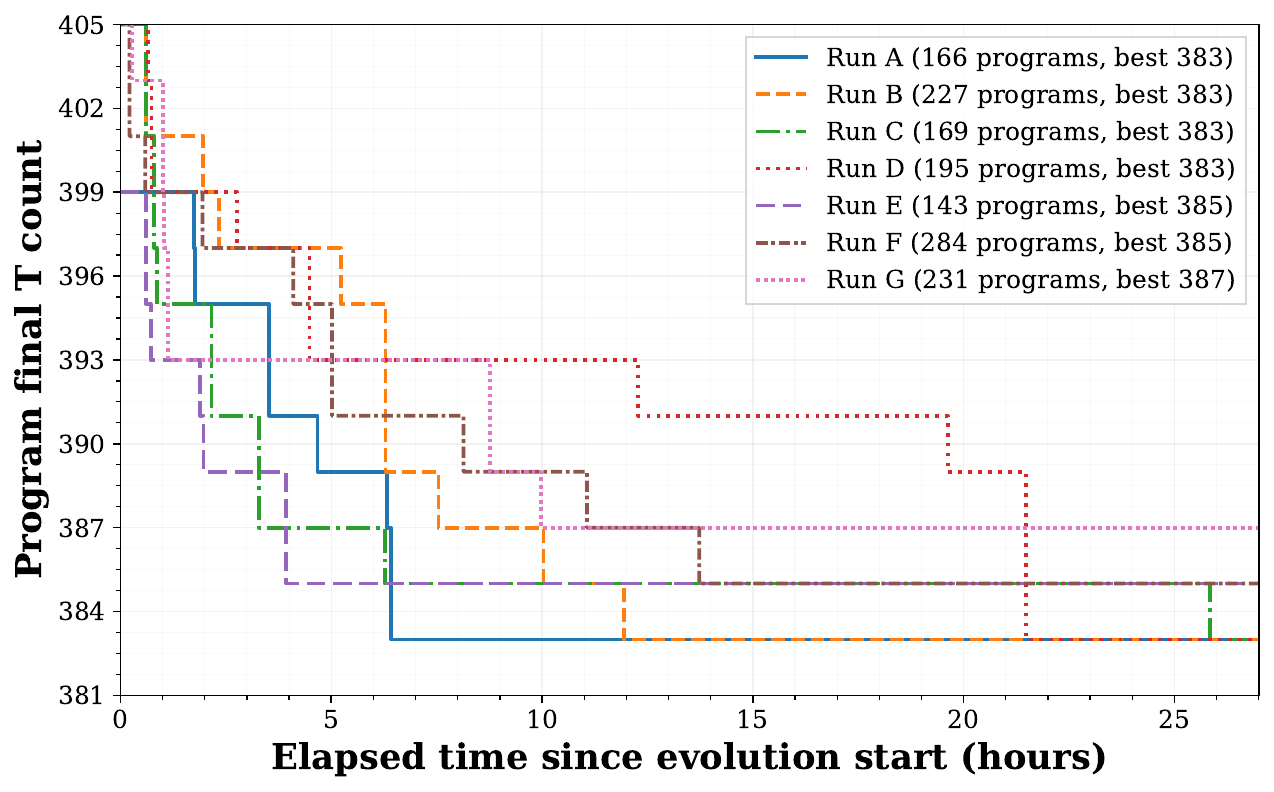}
    \caption{Best-so-far column count versus elapsed wall-clock time for separate VarTODD+\textsc{GigaEvo} outer evolution runs on $\mathrm{GF}(2^{16})$. Each curve tracks the incumbent best validated decomposition found by one run. The legend reports the number of evaluated tuner programs and the final best column count for each run. The runs differ in random seed and, in some cases, in the initial tuner program.}
    \label{fig:gigaevo_run_convergence}
\end{figure}

The diagnostic figures complement the benchmark tables by showing how the reported best-found $T$-counts emerge during outer program evolution. Figure~\ref{fig:gigaevo_cost_diagnostics} plots the results of evaluated tuner programs against cumulative LLM API cost for several $\mathrm{GF}(2^n)$ multiplication instances, whereas Fig.~\ref{fig:gigaevo_run_convergence} shows the corresponding best-so-far improvement as a function of elapsed wall-clock time for separate $\mathrm{GF}(2^{16})$ runs.

Many evaluated programs do not improve the incumbent, while occasional successful proposals produce additional downward steps in the best-found $T$-count. The rolling median also generally decreases over the course of the search, showing that the distribution of observed candidate outcomes shifts toward lower column counts. Thus, progress is not driven solely by isolated high-performing outliers: later stages of the evolutionary search tend to produce stronger tuner programs overall.

The seven separate $\mathrm{GF}(2^{16})$ runs are concentrated near the best observed result. Four runs reached a final $T$-count of $383$, two reached $385$, and one reached $387$. These outcomes indicate limited variation across runs for this instance under the tested configuration.

\section{Discussion}
\label{sec:discussion}

\subsection{Scope of the \texorpdfstring{$T$}{T}-count objective}

The preferred non-Clifford primitive depends on the target architecture, resource state factory, and workload. For some arithmetic problems, including the $\mathrm{GF}(2^n)$ multiplication circuits considered here, direct synthesis with Toffoli or CCZ gates may have a lower total cost than decomposing the circuit into Clifford+$T$ gates and subsequently minimizing its $T$-count. This can occur when Toffoli or CCZ resource states can be prepared and consumed efficiently, or when circuit depth, ancilla requirements, and scheduling constraints dominate the total cost~\cite{GidneyFowler2019CCZFactory,GidneyEkera2021RSA2048, khoruzhii2026tensordecompositionnoncliffordgate}.

Conversely, in architectures based primarily on $T$-state distillation and injection, the $T$-count provides a natural measure of aggregate non-Clifford resource demand~\cite{BravyiKitaev2005,Litinski2019MagicStateDistillation,Gosset2014AlgorithmTCount}. It is also the native optimization objective of the phase-polynomial and parity-matrix representation considered in this work, and many synthesis and optimization procedures act directly on Clifford+$T$ or phase-polynomial circuit components~\cite{Gosset2014AlgorithmTCount,todd,FastTodd}. The reductions reported here therefore translate directly into improved Clifford+$T$ realizations and, for $T$-state-based fault-tolerant architectures, reduced non-Clifford resource requirements while the precise impact on total physical cost remains architecture dependent.

\subsection{Program evolution and alternative approaches to policy search}

Reinforcement learning provides an alternative way to automate action selection. A learned policy can be trained across a distribution of circuit instances and, once trained, potentially transfer to new instances while amortizing the training cost. This approach, however, requires explicit choices for the state and action representations, reward, policy architecture, and training distribution, and its effectiveness depends on how well the training instances represent the problems encountered at inference time.

\textsc{GigaEvo} follows a different tradeoff. Rather than learning a single policy intended to generalize across instances, it performs an instance-specific search over executable tuner programs. A pretrained LLM proposes modifications to both numerical settings and search logic, allowing a tuner to combine different optimization procedures, schedules, and restart strategies without committing to a fixed neural policy architecture. This provides greater flexibility to adapt the search procedure to a particular circuit, at the cost of repeating the evolutionary search for each new instance. The two approaches therefore emphasize different forms of reuse: reinforcement learning aims to transfer a learned policy across instances, whereas program evolution reuses the capabilities of a pretrained model while adapting the search program to the instance at hand. The present results show that instance-specific program evolution can automate part of heuristic design for discrete, non-differentiable compiler searches.

\section{Conclusion}
\label{sec:conclusion}

We showed how a configurable extension of a fixed algebraic core can be explored with current AI methods, using $T$-count minimization with VarTODD+\textsc{GigaEvo} as an example. This approach exploits the ability of LLMs to control high-level design choices rather than leaving all such choices to a human designer, thereby helping to automate the search for heuristics tailored to a specific problem instance.

The VarTODD+\textsc{GigaEvo} pipeline matched or improved upon the lowest listed $T$-count for every evaluated benchmark, with a mean reduction of $7.0\%$ across the $\mathrm{GF}(2^n)$ multiplication circuits. For many instances, the resulting $T$-counts were below the corresponding FastTODD values, and for several they were also below the listed AlphaTensor Quantum RL values. Our repeated experiments on $\mathrm{GF}(2^{16})$ also showed limited variability across independent evolution runs: all seven runs finished within four $T$ gates of the best observed value, and four reached it. This indicates that the improvement for this instance did not depend on a single favorable evolutionary trajectory.

We believe that such an approach can be useful beyond $T$-count minimization. Many optimization algorithms combine a fixed mathematical core with high-level heuristic choices that are difficult to design and tune manually. By exposing these choices as a programmable search space, LLM-guided evolution can automate part of heuristic design while the underlying transformations and correctness conditions remain fixed and independently verifiable. The present results therefore suggest a practical way to use current AI systems to improve established optimization methods without asking them to replace the trusted algorithmic core.

\section*{Data availability}

The VarTODD implementation, including its Python bindings, benchmark decompositions, and scripts used to reproduce the reported results, is publicly available in the VarTODD repository~\cite{VarTODDRepository}. The open-source \textsc{GigaEvo} framework is available separately~\cite{GigaEvoRepository}, together with the experiment-specific fork used for the VarTODD+\textsc{GigaEvo} searches~\cite{GigaEvoVarTODDRepository}. The resulting parity matrices and validated $T$-count results reported in this work are included in the VarTODD repository.

\appendix
\section{Phase polynomials, parity matrices, and TODD-style updates}
\label{app:parity_background}

\subsection{\texorpdfstring{Phase-polynomial normal form for \{CNOT,$T$\} circuits}{Phase-polynomial normal form for CNOT+T circuits}}
\label{app:phase_polynomial_normal_form}

For a Hadamard-free \{CNOT,$T$\} circuit, or for the diagonal component obtained after gadgetization or circuit partitioning, the remaining non-Clifford content can be represented in phase-polynomial form~\cite{rm_optimizer,todd,FastTodd}. Concretely, any \{CNOT,$T$\} circuit on $n$ qubits induces a reversible linear map $A\in\mathrm{GL}(n,\mathbb{F}_2)$ and a function $f:\mathbb{F}_2^n\to\mathbb{Z}_8$ such that, on computational-basis states,
\begin{equation}
\label{eq:phase_poly_normal_form}
U\,|x\rangle \,=\, \omega^{\,f(x)}\,|Ax\rangle,
\qquad
\omega := e^{i\pi/4},
\qquad x\in\mathbb{F}_2^n.
\end{equation}
We refer to $f(x)$ as the \emph{phase polynomial}. The linear reversible component $A$ can be synthesized separately as a CNOT circuit; the present work focuses on the non-Clifford diagonal part.

The phase polynomial can be written as a sum of binary parities,
\begin{equation}
\label{eq:p_as_sum_of_parities}
f(x) \equiv \sum_{j=1}^{m} c_j \,\ell_j(x) \pmod{8},
\qquad
\ell_j(x) := \langle p_j, x\rangle_{\mathbb{F}_2},
\end{equation}
where each $p_j\in\mathbb{F}_2^n$ specifies a parity and $c_j\in\mathbb{Z}_8$ specifies its phase coefficient. Odd coefficients correspond to non-Clifford contributions.

In the multilinear expansion of a Boolean parity over $\mathbb{Z}_8$, every term of degree at least four has a coefficient divisible by 8 and therefore vanishes modulo $8$. Equivalently, the phase polynomial of a \{CNOT,$T$\} circuit has degree at most three in this representation. Consequently, third-order signature data are sufficient to capture the non-Clifford content relevant to $T$-count optimization.

Collecting the parity vectors with odd coefficients gives the \emph{parity matrix:}
\begin{equation}
\label{eq:parity_matrix_def}
P = \bigl[p_1\; p_2\; \cdots\; p_m\bigr] \in \mathbb{F}_2^{n\times m},
\end{equation}
where the number of columns $m$ is the current $T$-count.

The central compilation task in TODD-like optimizers is to replace $P$ by an \emph{equivalent} parity matrix with fewer columns. For $T$-count optimization, equivalence means preserving the non-Clifford phase data, with any remaining linear reversible and even-phase Clifford components tracked separately in the phase-polynomial representation.

This equivalence class is captured by the \emph{third-order symmetric signature tensor} $S(P)\in\mathbb{F}_2^{n\times n\times n}$, with repeated indices included, defined by
\begin{equation}
\label{eq:signature_tensor_appendix}
S_{\alpha,\beta,\gamma}(P)
\equiv
\bigl|P_\alpha \wedge P_\beta \wedge P_\gamma\bigr|
\pmod 2,
\end{equation}
where $P_\alpha$ denotes row $\alpha$ of $P$, $\wedge$ is bitwise AND, and $|\cdot|$ is Hamming weight. Equivalently,
\begin{equation}
\label{eq:signature_tensor_sum_form}
S_{\alpha,\beta,\gamma}(P)
\equiv
\sum_{j=1}^{m} P_{\alpha,j} P_{\beta,j} P_{\gamma,j} \pmod 2.
\end{equation}
Two parity matrices are equivalent, up to diagonal Clifford corrections, if and only if they induce the same signature tensor~\cite{todd,rm_optimizer,FastTodd,Ruiz2025AlphaTensorQuantum}. Therefore, minimizing $T$-count becomes the problem of finding a parity matrix $P^{\prime}$ with $S(P^{\prime})=S(P)$ and a minimal column count.

\subsection{TODD/FastTODD update rule and admissible actions}
\label{app:fasttodd_details}

We interpret the search as an iterative process over parity matrices with a large action space, following TODD-style updates~\cite{todd,FastTodd}. An action is a pair of binary vectors $(z,y)$ with $z\in\mathbb{F}_2^n$ and $y\in\mathbb{F}_2^m$; applying an action first forms the intermediate matrix
\begin{equation}
\label{eq:todd_update_appendix}
\widetilde P = P \oplus z y^{\mathsf T},
\end{equation}
where $\oplus$ is bitwise XOR\@. TODD/FastTODD impose linear constraints on $y$ (for fixed $z$ and current $P$) so that this transformation preserves the signature tensor. For a fixed $z$, the set of admissible vectors forms a linear space
\[
\mathcal{N}_z(P) \subseteq \mathbb{F}_2^m,
\]
and FastTODD computes generators for these null spaces efficiently before searching for actions that maximize cancellation~\cite{FastTodd}.

After an admissible update, the algorithm applies the simplification
\begin{equation}
\label{eq:odd_update_appendix}
P^{\prime} = \operatorname{odd}(\widetilde P),
\end{equation}
where $\operatorname{odd}(\cdot)$ keeps precisely the columns that occur with odd multiplicity, equivalently removing all-zero columns and canceling duplicate columns in pairs. These simplifications preserve the same signature tensor and only modify the diagonal Clifford part, which can always be restored from the phase-polynomial representation. Thus, $P^{\prime}$ remains equivalent to $P$ while potentially having fewer columns.

A practical observation is that reduction in column count requires creating duplicate or zero columns after the XOR update by $z$. Accordingly, candidate $z$ values are typically drawn from a structured set derived from the current columns:
\begin{equation}
\label{eq:z_candidates_appendix}
\mathcal{Z} = \{P_{:,i}\oplus P_{:,j} \mid 1\le i<j\le m\}
\cup
\{P_{:,i} \mid 1\le i\le m\}.
\end{equation}
In the worst case, $|\mathcal{Z}|=\mathcal{O}(m^2)$. The original TODD algorithm may require expensive linear algebra per $z$ and has worst-case complexity $\mathcal{O}(n^3m^5)$~\cite{todd}. FastTODD reorganizes the feasibility checks and reduces the worst-case complexity to $\mathcal{O}(n^4m^3)$, while often improving solution quality in practice~\cite{FastTodd}.

\subsection{TOHPE as a lightweight exploration stage}
\label{app:tohpe_details}

For early-stage exploration, a lighter TOHPE procedure can be used~\cite{FastTodd}. TOHPE computes a common subspace
\begin{equation}
\label{eq:tohpe_space_appendix}
\mathcal{N}_{\mathrm{tohpe}} = \bigcap_{z\in\mathbb{F}_2^n} \mathcal{N}_z(P),
\end{equation}
which can be constructed with complexity $\mathcal{O}(nm^2)$~\cite{FastTodd}. This enables inexpensive generation of candidate actions and can be used to identify promising $z$ values before invoking more expensive FastTODD-style null space computations. In practice, TOHPE exploration is useful as a screening stage: it can populate an action pool cheaply and guide subsequent deeper search in the larger admissible spaces $\mathcal{N}_z(P)$.

\section{VarTODD implementation details}
\label{app:vartodd}

This appendix specifies the VarTODD policy interface and search controls used in the combined VarTODD+\textsc{GigaEvo} experiments; low-level details of the C++/Python bindings are available in the public implementation. The algebraic objects are those introduced in Appendix~\ref{app:parity_background}. At each iteration, the current state is a parity matrix \(P\), and an admissible action is a pair \(a=(z,y)\) whose update preserves the third-order signature tensor. VarTODD leaves the admissibility conditions unchanged; tunability enters through the policy controlling candidate generation, retention, scoring, selection, scheduling and branching.

Throughout this appendix, the \(T\)-count is the number of columns remaining in
the parity matrix after zero columns are removed and duplicate columns are
canceled. Accordingly, a \(T\)-count reduction refers to the reduction measured
after these simplifications.

\subsection{Candidate sources and action pools}
\label{app:search_controls}

A VarTODD iteration constructs a finite set of admissible candidate actions from
one or more candidate sources. In the implementation used here, the two main
sources are:

\begin{description}[leftmargin=2.9cm,style=nextline]
    \item[TOHPE source]
    This source generates candidate actions from the common TOHPE admissible
    space by sampling \(y\) and retaining the \(k_{\mathrm{tohpe}}\) candidate
    values of \(z\) that yield the greatest immediate \(T\)-count reduction.

    \item[TODD/FastTODD source]
    This source traverses \(k_z\) candidate values of \(z\), constructs the
    corresponding null spaces \(\mathcal{N}_z(P)\), and samples or enumerates
    vectors \(y\) in those spaces subject to the implementation budget. It then
    forms admissible actions \(a=(z,y)\) and evaluates their resulting
    \(T\)-count reductions.
\end{description}

The retained candidates are merged into a final action pool, which is passed to
the final scoring and selection rule. The final-pool size controls a practical
tradeoff: a very large pool increases the cost of final scoring and comparison,
whereas a very small pool may discard useful candidates before they reach the final selection stage.

The main conceptual policy components are as follows:

\begin{description}[leftmargin=3.3cm,style=nextline]
    \item[Candidate-source budgets]
    Allocate work between inexpensive TOHPE-style exploration and deeper TODD
    search through the \(k_z\), \(k_{\mathrm{tohpe}}\), and \(y\)-sampling
    parameters.

    \item[Source retention]
    Limit how many candidates from each source, or from each
    \(\mathcal{N}_z(P)\) space, may enter the merged action pool.

    \item[Final-pool size]
    Control how many generated candidates are considered by the final
    selection rule.

    \item[Scoring weights]
    Rank candidates using features such as immediate \(T\)-count reduction,
    null-space dimension, bucket potential, and Hamming-weight information.

    \item[Action-selection temperature]
    Interpolate between an approximately greedy choice and more exploratory
    stochastic selection among the retained candidates.

    \item[Beam or multi-action width]
    Allow several successor states or selected actions to be explored
    simultaneously, increasing robustness at additional computational cost.

    \item[\(T\)-count schedules]
    Use different policy settings in states with a high \(T\)-count and in the
    more difficult terminal regime with a low \(T\)-count.
\end{description}

\subsection{Choice-stage scoring and final selection}
\label{app:choice_scoring}

The final selection rule scores only actions that have already been generated
and retained. Its features include the immediate \(T\)-count reduction, the
dimension of the relevant admissible space, a measure of the potential of the
selected \(z\) bucket, Hamming-weight information for \(y\) and \(z\), and, in
some policies, a lightweight lookahead feature. A convenient family of scores
used by the evolved tuner programs is

\begin{equation}
    s(a)=\sum_i w_i\,\lvert u_i(a)-c_i\rvert^p,
\label{eq:score_appendix}
\end{equation}

where \(u_i(a)\) are normalized candidate features, \(w_i\) are weights,
\(c_i\) are centers, and \(p\) is the score exponent. Candidates with higher scores are preferred.
With zero centers, positive weights favor larger feature values, whereas negative
weights penalize them.
Nonzero centers allow the policy to prefer intermediate feature values, such as
a particular range of null-space dimensions.

\subsection{\texorpdfstring{\(T\)-count}{T-count} schedules and validation}
\label{app:rank_schedules_details}

Most policy components may either remain constant throughout a run or vary with
the current \(T\)-count \(\rho\). A typical piecewise-constant schedule has the
form

\begin{equation}
\theta(\rho)=
\begin{cases}
\theta_0, & \rho \ge r_1,\\
\theta_i, & r_{i+1}\le \rho < r_i,\\
\theta_k, & \rho < r_k,
\end{cases}
\label{eq:schedule_appendix}
\end{equation}

where \(r_1 > r_2 > \cdots > r_k\) are \(T\)-count thresholds and the middle
case applies for \(1 \le i \le k-1\). Such schedules allow a policy to use
broad, inexpensive exploration at a high \(T\)-count and more selective or
deeper search near the end of the reduction trajectory.

After each selected action is applied, zero columns are removed and duplicate
columns are canceled as in Eq.~\eqref{eq:odd_update_appendix}. The evaluator
then records the resulting \(T\)-count and the associated diagnostic
statistics. Final results reported in this work are accepted only after an external evaluator independently confirms
\begin{equation}
S(P_{\mathrm{final}})=S(P_0).
\label{eq:appendix_external_validation}
\end{equation}

\definecolor{lstBlockGreen}{RGB}{20,90,60}
\definecolor{lstInlineGreen}{RGB}{38,105,72}
\definecolor{lstRuleGray}{RGB}{120,120,120}

\lstdefinestyle{vtpycompact}{
  language=Python,
  basicstyle=\ttfamily\fontsize{7.2pt}{8.35pt}\selectfont,
  keywordstyle=\bfseries\color{black!85},
  stringstyle=\color{black!85},
  commentstyle=\ttfamily\itshape\color{lstInlineGreen},
  morecomment=[s][\sffamily\color{lstBlockGreen}]{/*}{*/},
  linewidth=\columnwidth,
  columns=fullflexible,
  keepspaces=true,
  showstringspaces=false,
  numbers=none,
  tabsize=4,
  breaklines=true,
  breakatwhitespace=true,
  breakautoindent=true,
  breakindent=0.9em,
  frame=tb,
  rulecolor=\color{lstRuleGray},
  framerule=0.35pt,
  framesep=2pt,
  xleftmargin=0pt,
  xrightmargin=0pt,
  aboveskip=0.45\baselineskip,
  belowskip=0.45\baselineskip,
  abovecaptionskip=0.35\baselineskip,
  belowcaptionskip=0pt,
  captionpos=b
}
\section{Evolutionary trace of the
\texorpdfstring{$\mathrm{GF}(2^{32})$}{GF(2^32)} tuner}
\label{app:gf32_evolution_trace}
The run snapshot contains 368 tuner programs across 14 ancestry generations.
Of 357 completed evaluations, 286 returned valid decompositions. The best
validated \(T\)-count among the eight initial programs was 1205; the first program to reach 1177 was at ancestry generation 11. Because evaluations were concurrent and the evolutionary process was steady-state, generation records lineage depth rather than wall-clock order.

\subsection{Best tuner in the initial population}
\label{app:gf32_initial_programs}

The best initial tuner reached \(T=1205\) from the input parity matrix. It
used an 80-evaluation DE scout with the TOHPE source, reopened the resulting
trajectory, enabled TODD/FastTODD, and ran a 120-evaluation PatternSearch
tail. Listing~\ref{lst:gf32_best_initial} retains the corresponding search
settings while omitting parameter-mapping and evaluator boilerplate.

\begin{lstlisting}[
  style=vtpycompact,
  caption={Condensed API-level pseudocode for the best initial tuner
  (initial population; \(T\)-count 1205).},
  label={lst:gf32_best_initial}
]
SCOUT_EVALS = 80
TAIL_EVALS = 120
REOPEN_MARGIN = 58

# All score weights are mapped independently over [-4, 4].
set_scores(
    PolicyScores(
        exploration=ExplorationScore(
            weights=mapped_vector(5, -4, 4), centers=zeros(5), pow=1
        ),
        final=FinalizationScore(
            weights=mapped_vector(6, -4, 4), centers=zeros(6), pow=1
        ),
    )
)

# The inexpensive scout constructs an ab-initio TOHPE trajectory.
set_action_selection(
    ActionSelection(beamwidth=1, mode="softmax", temperature=0.30)
)
set_action_pool(ActionPool(final_size=mapped_int(12, 26)))
set_tohpe_search(
    TohpeSearch(
        sampling=SamplingBudget(
            one_hot="all", sparse=0, dense=6,
            sparse_max_weight=2,
        ),
        pool=SourcePool(
            keep=mapped_int(8, 24), reserve=mapped_int(1, 5)
        ),
        z_choices=mapped_int(4, 14),
    )
)
set_todd_search(enable=False)

params = minimize_from_init(
    DE(pop_size=8, CR=0.85, F=0.65), n_eval=SCOUT_EVALS
)

# The tail reopens the scout trajectory and admits TODD/FastTODD.
set_action_selection(
    ActionSelection(beamwidth=2, mode="softmax", temperature=0.14)
)
set_action_pool(ActionPool(final_size=mapped_int(20, 40)))
set_tohpe_search(
    TohpeSearch(
        sampling=SamplingBudget(
            one_hot="all", sparse=6, dense=10,
            sparse_max_weight=3,
        ),
        pool=SourcePool(keep=8, reserve=2),
        z_choices=5,
    )
)

set_todd_search(
    ToddSearch(
        sampling=SamplingBudget(
            one_hot="all", sparse=10, dense=10,
            sparse_max_weight=3,
        ),
        pool=SourcePool(keep=mapped_int(3, 6), reserve=1),
        actions_per_bucket=mapped_int(2, 3),
        buckets=ZBucketSearch(
            min_buckets=1200,
            max_buckets=300_000,
            limit_bucket=300_000,
        ),
    )
)

path = reopen_best_trajectory(
    path_num=0, margin=REOPEN_MARGIN, xopt=params
)
minimize_from_loaded_path(
    PatternSearch(x0=params, init_delta=0.2),
    path, n_eval=TAIL_EVALS
)
\end{lstlisting}

\subsection{Recorded evolution statistics}
\label{app:gf32_evolution_evidence}

The run produced 360 non-root programs through three recorded mutation
routes. Table~\ref{tab:gf32_evolution_evidence} reports their observed
evaluation and outcome statistics.

\begin{table*}[t!]
\centering
\caption{Observed outcomes by mutation route. Percentages in the mutation
column use all 360 non-root programs; validity percentages use all programs
within the corresponding route.}
\label{tab:gf32_evolution_evidence}
\footnotesize
\setlength{\tabcolsep}{3.5pt}
\begin{tabular}{lrrrrrr}
\hline
\shortstack{Mutation\\route} &
\shortstack{Mutations\\\(n\) (\%)} &
\shortstack{Valid\\\(n\) (\%)} &
\shortstack{Final \(T\)-count\\median [IQR]} &
\shortstack{\(T\leq1179\)\\among valid} &
\shortstack{Strict cached-path\\improvements} &
\shortstack{Best\\\(T\)-count} \\
\hline
Ab-initio  & 118/360 (32.8\%) & 97/118 (82.2\%) & 1209 [1203, 1215] & 0/97            & --               & 1193 \\
Mid-margin & 118/360 (32.8\%) & 86/118 (72.9\%) & 1195 [1185, 1201] & 10/86 (11.6\%) & 28/86 (32.6\%)  & 1177 \\
Near-end   & 124/360 (34.4\%) & 95/124 (76.6\%) & 1187 [1181, 1201] & 20/95 (21.1\%) & 26/92 (28.3\%)  & 1177 \\
\hline
\end{tabular}
\end{table*}

Ab-initio programs start from the input matrix; mid-margin and near-end
programs reopen cached trajectories at different branch depths. A strict
cached-path improvement ends below the final \(T\)-count of the loaded path.
Its denominator includes only valid programs with a recorded loaded path.
The routes used different starting paths and ancestry generations, and many
programs changed several settings together; the table therefore reports
observed groups rather than isolated parameter comparisons.

The successive run incumbents were
\(1205,1201,1195,1193,1183,1181,1179,1177\), first recorded at ancestry
generations \(1,2,4,5,7,9,10,11\), respectively.

\subsection{First tuner to reach \texorpdfstring{\(T=1177\)}{T=1177}}
\label{app:gf32_best_final_program}

Listing~\ref{lst:gf32_best_final_program} shows the first tuner to reach
\(T=1177\). It loaded a separately constructed trajectory ending at 1183,
scheduled TOHPE sampling according to the current $T$-count, used deterministic beam
selection, and ran DE from the loaded trajectory and reopened branches. The
database-specific path handle is replaced by a descriptive reference.

\begin{lstlisting}[
  style=vtpycompact,
  caption={Condensed API-level pseudocode for the first tuner to reach
  \(T=1177\) (ancestry generation 11).},
  label={lst:gf32_best_final_program}
]
TAIL_EVALS = 100
N_RESTARTS = 3

EVALUATION_SEEDS = (23,)
T_COUNT_SPAN = max(1, INITIAL_T_COUNT - TARGET_FINAL_T_COUNT)
# G10 first used this margin/reopen rule: the loaded T=1183 path
# reopened at T=1224 and reached T=1179 at evaluation 40/213.
MARGIN = 40
# In G11, the value 49 reopened at T=1234; T=1177 first appeared
# at evaluation 1659/1758
REOPEN_MARGIN = 49
TERMINAL_TCOUNT = 1210
HARD_BUCKET_GUARD = 5000

set_scores(
    PolicyScores(
        exploration=ExplorationScore(
            weights=[
                mapped(0.3, 3.0), mapped(0.3, 2.0),
                mapped(-4.0, 2.0), mapped(-4.0, 0.0),
                mapped(-2.0, 4.0),
            ],
            centers=zeros(5), pow=1,
        ),
        final=FinalizationScore(
            weights=[
                mapped(0.0, 4.0), mapped(-2.0, 4.0),
                mapped(-4.0, 2.0), mapped(-4.0, 0.0),
                mapped(-2.0, 4.0), mapped(0.0, 4.0),
            ],
            centers=zeros(6), pow=1,
        ),
    )
)

# G7 first used the T=1210 split. Returned-path mean dimension was
# 20.8 at T=1234-1210, 8.0 at 1210-1182, and 2.7 at 1182-1177.
high_tcount_tohpe = TohpeSearch(
    sampling=SamplingBudget(
        one_hot="all",
        sparse=mapped_int(4, 10),
        dense=mapped_int(6, 14),
        sparse_max_weight=3,
    ),
    pool=SourcePool(
        keep=mapped_int(14, 30), reserve=mapped_int(2, 6)
    ),
    z_choices=mapped_int(8, 18),
)
# G10 first paired this sparse=dense=1 budget with the T-count split;
# at dimension 2--3, the nonzero y-space has only 3--7 vectors.
terminal_tohpe = TohpeSearch(
    sampling=SamplingBudget(
        one_hot="all", sparse=1, dense=1,
        sparse_max_weight=3,
    ),
    pool=SourcePool(keep=12, reserve=3),
    z_choices=8,
)
set_tohpe_search(
    t_counts=[TERMINAL_TCOUNT],
    values=[high_tcount_tohpe, terminal_tohpe],
)

# G11 first used the 5000-bucket guard: 213 -> 1758 evaluations
# in about 6224 s; the returned T=1177 path used 2126 buckets.
set_todd_search(
    ToddSearch(
        sampling=SamplingBudget(
            one_hot=6, sparse=2, dense=2,
            sparse_max_weight=2,
        ),
        pool=SourcePool(
            keep=mapped_int(60, 80),
            reserve=mapped_int(20, 30),
        ),
        actions_per_bucket=mapped_int(60, 100),
        buckets=ZBucketSearch(
            min_buckets=1024,
            max_buckets=min(30000, buckets_space),
            limit_bucket=HARD_BUCKET_GUARD,
        ),
    )
)

# G10 first combined beam 12--16 and pool 60--80; the observed terminal pool grew from 21 to 43 actions. "best" first appeared G6; beam=14 here.
set_action_pool(ActionPool(final_size=mapped_int(60, 80)))
set_action_selection(
    ActionSelection(
        beamwidth=mapped_int(12, 16), mode="best"
    )
)

# G9 first used this DE setup, three restarts, and alternating branches; its introducing tuner ended at T=1183. One evaluator seed appeared in G1.
path = cached_trajectory(phase="middle", final_t_count=1183)
policy = minimize_from_loaded_path(
    DE(pop_size=20, CR=0.98, F=0.60),
    path, margin=MARGIN, n_eval=TAIL_EVALS, seed=23,
)

for restart, branch in alternating_restarts(N_RESTARTS):
    path = reopen_saved_branch(
        branch,
        t_count_margin=REOPEN_MARGIN,
        xopt=policy if restart == 0 else None,
    )
    policy = minimize_from_loaded_path(
        DE(pop_size=20, CR=0.98, F=0.60),
        path, n_eval=TAIL_EVALS,
        seed=23 + 7 * restart,
    )

return best_trajectory()
\end{lstlisting}

\medskip
\bibliography{biblio}

\end{document}